\documentstyle[b98proc]{article}

\def\Journal#1#2#3#4{{#1} {\bf #2}, #3 (#4)}

\def\NPA{{\em Nucl. Phys.} A}

\def\PLB{{\em Phys. Lett.}  B}

\def\PRC{{\em Phys. Rev.} C}

\def\PREP{\em Phys. Rep.}

\begin{document}

\title{IMPORTANT CONFIGURATIONS FOR $NN$ PROCESSES IN A GOLDSTONE 
BOSON EXCHANGE MODEL}

\author{ D. BARTZ and FL. STANCU }

\address{Institute of Physics, B.5, University of Li\`ege, 
Sart Tilman, B-4000 Li\`ege 1, Belgium}

\maketitle\abstracts{
We study the short-range nucleon-nucleon interaction in a nonrelativistic
chiral constituent quark model by diagonalizing a Hamiltonian
containing a linear confinement and a Goldstone boson exchange
interaction between quarks. A finite six-quark basis obtained
from single particle cluster model states
was previously used. Here
we show that the configurations which appear naturally through the use of
molecular orbitals, instead of cluster model states,
are much more efficient in lowering the six-quark
energy.}

Constituent quark models have been applied to the study of the nucleon-nucleon
interaction. The Hamiltonian of such models usually contains a kinetic term, a
confinement term and an effective one-gluon exchange (OGE) term. These models
explain the short-range repulsion in the $NN$ systems as due to the
colour-magnetic part of the OGE interaction combined with quark interchanges
between the $3q$ clusters. Nevertheless, an effective meson-exchange potential,
introduced through the coupling of mesons to $3q$ cluster collectively,
is required in order to reproduce the intermediate- and long-range attraction.
Another category are the hybrid models.
There, in addition to OGE interaction, the quarks belonging to different $3q$
clusters interact via pseudoscalar and scalar meson exchange. In these models
the short-range repulsion in the $NN$ system is still attributed to the OGE
interaction between the constituent quarks. The medium- and the long-range
attraction are due to meson-exchange, as expected. \par
In a recent exploratory work \cite{ST97}, by using the Born-Oppenheimer
approximation, we calculated an effective $NN$ interaction at zero separation
distance, within the constituent quark model \cite{GL96a,GL96b,GL97}. In
this model the quarks interact via Goldstone boson exchange (GBE)
instead of OGE of conventional models, and the
hyperfine splitting in hadrons is obtained from the short-range part of the GBE
interaction. An important merit of the GBE model is that it reproduces the
correct order of positive and negative parity states in both nonstrange
\cite{GL96b} and strange baryons \cite{GL97}. In Ref. \cite{ST97} we showed
that the same short-range part of the GBE interaction, also induces a
short-range repulsion in the $NN$ system.\par
In Ref. \cite{ST97} the
height of the repulsive core was about 800 MeV for the $^3S_1$ channel and
1300 MeV for the $^1S_0$ channel. Such a result has been obtained from
diagonalizing the Hamiltonian of Ref. \cite{GL96b} in a six-quark
cluster model basis built from harmonic oscillator states
containing up to two quanta of excitation.
The six-quark states have orbital symmetries $[6]_O$ and $[42]_O$,
so that they contain configurations
of type $s^6$, $s^4p^2$
and $s^52s$, with the centre of mass motion removed.
In the flavour-spin space only
the symmetries $[33]$, $[51]$ and $[411]$ were retained.
As shown in \cite{ST97} they produce the most important
five basis states allowed by the Pauli principle. Due
to the specific flavour-spin structure of the GBE interaction, we found
that the
state $|s^4p^2[42]_O[51]_{FS}\rangle$ was highly dominant at zero-separation
between nucleons. The symmetry structure of this state implies the existence of
a node in the nucleon-nucleon
$S$-wave relative motion wave function at short distances.
This nodal structure will induce an additional effective repulsion
in dynamical calculations based , for example, on the resonating
group method.\par
A central issue of the $NN$ problem is the construction of an adequate
six-quark
basis states. In principle the choice of basis is arbitrary if a sufficiently
large basis is considered in the Hamiltonian diagonalization. But, as in
practice one considers a finite set, its choice is very important. Ref.
\cite{ST87} advocated the use of molecular-type single particle orbitals
instead of cluster model-type states. These orbitals have the proper axially
and reflectionally symmetries and can be constructed from appropriate
combinations of two-centre Gaussians. At zero-separation the six-quark states
obtained from such orbitals contain certain $p^ns^{6-n}$ components which are
missing in the cluster model basis. In Ref. \cite{ST88} it has been shown
that for an OGE model used in the calculations of the $NN$ potential
they lead to a substantial lowering of the lowest eigenstate,
used in the calculation of the $NN$ potential. The molecular orbitals have also
the advantage of forming an orthogonal and complete basis while the cluster
model (two-centre) states are not orthogonal and are overcomplete.\par
Due to the predominance ( 93 \%) of only one component,
$|s^4p^2[42]_O[51]_{FS}\rangle$, in the ground state wave function obtained
in a
cluster model basis \cite{ST97} the GBE model is a more chalenging case to
test the efficiency of a molecular orbital basis than the OGE model, where
there is some mixture of states (see e.g. \cite{ST88}). Here we show that
by using molecular orbitals the height of the repulsion reduces by about
22 \% and 25 \% in the $^3S_1$ and $^1S_0$
channels respectively. \par

To this end we diagonalize the Hamiltonian of Ref. \cite{GL96b} 
in a six-quark basis 
constructed 
from single particle molecular-type orbitals defined as linear combinations
of the same $s$ and $p$ states, as used in the cluster model 
study \cite{ST97}. We calculate
the $NN$ interaction potential in the Born-Oppenheimer approximation
\begin{equation}
V_{NN}\left(Z\right) = \langle H\rangle_Z - \langle H\rangle_{\infty}\ ,
\end{equation}
where  $\langle H\rangle_Z$ is the lowest expectation value
obtained from the diagonalization at a given $Z$ and $\langle H\rangle_{\infty}
= 2m_N$ is the energy (mass) of two well separated nucleons. Here we
study the case $Z = 0$, relevant for short separation distances between the
nucleons. In Tables I and II we present our results for $IS$ = (01) and (10)
respectively, obtained from the diagonalization of $H$. From the diagonal
matrix elements $H_{ii}$ as well as from the eigenvalues, the quantity $2m_N$ =
1939 MeV has been subtracted. Here $m_N$ is the nucleon mass
calculated also variationally, with an $s^3$ configuration.
This value is
obtained for a harmonic oscillator parameter $\beta$ = 0.437 fm.
For sake of comparison with Ref. \cite{ST97} we take the same 
$\beta$ for the six-quark system as well.\par
In both $IS$=(01) and (10) cases the effect of using
molecular orbitals is rather remarkable in lowering the ground state
energy as compared to the cluster model value \cite{ST97}. 
Accordingly, the height of
the repulsive core in the $^1S_3$ channel
is reduced from 915 MeV in the cluster model basis 
to 718 MeV in the molecular orbital basis. In the $^1S_0$ channel
the reduction is from 1453 MeV to 1083 MeV.
Thus the molecular
orbital basis is much better, inasmuch as the same two single
particle states, $s$ and $p$, are used in both bases.

\vspace{-0.2cm}
\begin{table}[h]
{\caption[states]{\label{states} Results of the diagonalization of the
Hamiltonian \cite{GL96b} for
$IS$ = (01). Column 1 - basis states, column 2 - diagonal matrix elements (GeV),
column 3 - eigenvalues (GeV) in increasing order, column 4 - lowest state
amplitudes of components given in column 1}} 
\begin{tabular}{|l|c|c|c|}
\hline
\ \ \ \ \ \ \ \ \ State &$H_{ii}$ - 2 $m_N\,$ & Eigenvalues - 2 $m_N$ & Lowest state\\
& & & amplitudes\\
\hline
$|33[6]_O[33]_{FS} >$ & 2.616 & 0.718 & -0.04571\\
\hline
$|33[42]_O[33]_{FS} >$ & 3.778 & 1.667 & 0.02479\\
\hline
$|33[42]_O[51]_{FS} >$ & 1.615 & 1.784 & -0.31762\\
\hline
$|33[42]_O[411]_{FS} >$ & 2.797 & 2.309 & 0.04274\\
\hline
$|42^+[6]_O[33]_{FS} >$ & 3.062 & 2.742 & -0.07988\\
\hline
$|42^+[42]_O[33]_{FS} >$ & 2.433 & 2.784 & 0.12930\\
\hline
$|42^+[42]_O[51]_{FS} >$ & 0.850 & 3.500 & -0.93336\\
\hline
$|42^+[42]_O[411]_{FS} >$ & 3.665 & 3.752 & 0.00145\\
\hline
$|51^+[6]_O[33]_{FS} >$ & 2.910 & 4.470 & -0.01789\\
\hline
\end{tabular}
\end{table}


The previous study \cite{ST97},
performed in a cluster model basis indicated that the dominant configuration is
associated to the symmetry $[42]_O[51]_{FS}$. It is the case here too
and one can see from Tables I and II that
the diagonal matrix element $H_{ii}$ of the state
$|42^+[42]_O[51]_{FS} >$ is far the lowest one, so that this
state is
much more favoured
than $|33[42]_O[51]_{FS} >$ . Such a
state represents a configuration with
two quarks on the left and four on the right
around the symmetry centre.
At $Z \rightarrow \infty$ its energy becomes infinite and it does no more  
contribute to the ground state, i.e. it behaves as a hidden colour state. 
But at $Z = 0$ it is the dominant component
of the lowest state  with a probability of
87 \% for $IS$ = (01) and 93 \% for $IS$ = (10). The next
important state is
$|33[42]_O[51]_{FS} >$ with a probability of 10 \%
for $IS$ = (01) and 4 \% for $IS$ = (10).
The presence of this state will become more and more important
with increasing $Z$. Asymptotically this state corresponds to
a cluster model state with three quarks on the left and three
on the right of the symmetry centre.\par
Details of this study can be found in Ref. \cite{BS98}.
The following step will be to calculate the $NN$ potential at $Z \neq 0$.
The Yukawa potential tail, already contained in the GBE
interaction \cite{GL96b}  will bring
the required long-range attraction. It would be interesting to
find out the amount of middle-range attraction brought in by
two correlated or uncorrelated pion exchanges.

\vspace{-0.2cm}
\begin{table}[h]
{\caption[statesbis]{\label{statesbis} Same as Table I but for $IS$ = (10)}}
\begin{tabular}{|l|c|c|c|}
\hline
\ \ \ \ \ \ \ \ \ State &$H_{ii}$ - 2 $m_N$\,& Eigenvalues - 2 $m_N$ & Lowest state \\
& & & amplitudes\\
\hline
$|33[6]_O[33]_{FS} >$ & 3.300 & 1.083 & -0.02976\\
\hline
$|33[42]_O[33]_{FS} >$ & 4.367 & 2.252 & 0.01846\\
\hline
$|33[42]_O[51]_{FS} >$ & 2.278 & 2.279 & -0.20460\\
\hline
$|33[42]_O[411]_{FS} >$ & 3.191 & 2.945 & -0.04729\\
\hline
$|42^+[6]_O[33]_{FS} >$ & 3.655 & 3.198 & -0.07215\\
\hline
$|42^+[42]_O[33]_{FS} >$ & 2.796 & 3.317 & 0.13207\\
\hline
$|42^+[42]_O[51]_{FS} >$ & 1.167 & 4.058 & -0.96531\\
\hline
$|42^+[42]_O[411]_{FS} >$ & 4.405 & 4.459 & -0.00081\\
\hline
$|51^+[6]_O[33]_{FS} >$ & 3.501 & 5.070 & -0.01416\\
\hline
\end{tabular}
\end{table}


\end{document}